\documentclass[prl,twocolumn]{revtex4}

\usepackage{amsfonts}
\usepackage{amssymb}
\usepackage{graphicx}
\usepackage{pstricks}

\newcommand{\be}{\begin{equation}}
\newcommand{\bea}{\begin{eqnarray}}
\newcommand{\ee}{\end{equation}}
\newcommand{\eea}{\end{eqnarray}}
\newcommand{\bpi}{\begin{picture}}
\newcommand{\bce}{\begin{center}}
\newcommand{\epi}{\end{picture}}
\newcommand{\ece}{\end{center}}
\newcommand{\re}{{\mathrm Re}}

\begin{document}

\title{CPT Violating Decoherence and LSND: a possible window to Planck scale Physics}
\date{\today}

\author{Gabriela Barenboim$^a$}
\author{Nick E. Mavromatos$^{b}$}
\affiliation{$^a$Departamento de F\'\i sica Te\'orica and IFIC, Centro Mixto, 
Universidad de Valencia-CSIC,
E-46100, Burjassot, Valencia, Spain. \\
$^b$King's College London, University of London, Department of Physics, 
Strand WC2R 2LS, London, U.K.}

\begin{abstract}
Decoherence has the potential to explain all 
existing neutrino data including LSND results, 
without enlarging the neutrino sector.
This particular form of CPT violation can preserve the 
equality of masses and
mixing angles between particle and antiparticle sectors, and still provide 
seizable differences in the oscillation patterns.
A simplified minimal model of decoherence 
can explain the existing
neutrino data as well as the standard three oscillation scenario, 
while making dramatic predictions for
the upcoming experiments. Such a model can easily accomodate
the LSND result but cannot fit the spectral distortions seen by KamLAND. 
Some comments on the 
order of the decoherence parameters in connection with 
theoretically expected values 
from some models of quantum-gravity are given.
In particular, the quantum gravity decoherence as
a primary origin of the neutrino mass differences scenario is explored, and  
even a speculative link
between the neutrino mass-difference scale and the dark energy density component
of the Universe today is drawn. 
\end{abstract}



\maketitle

\vspace{1.cm}

\section{Introduction: The LSND puzzle and abandoning of Symmetries} 

Although the Special Theory of Relativity,
a theory of flat space time physics based on Lorentz symmetry, is 
very well tested, and in fact next year it celebrates 
a century of enormous success, 
having passed very stringent experimental precision tests, 
a quantum theory of Gravity,
that is a consistent quantized version of Einstein's General Relativity,
a dynamical model for curved space times, 
still eludes us.  
The main reason for this is the lack of any concrete observational
evidence on the structure of space time at the characteristic scale of 
quantum gravity (QG), the (four-dimensional) 
Planck mass scale $M_P = 10^{19}$ GeV. 
The situation concerning QG 
is to be contrasted with that
characterizing its classical counterpart, 
General 
Relativity, and the quantum version of Special Relativistic
field theories, both of which can be 
tested to enormous precision to date. 
For instance, the perihelion precession
of planets such as mercury, as well as the deflection of light by the sun,
made the General Theory of Relativity an instant success, almost
immediately after it was proposed in 1915. Similarly, the modern 
versions of relativistic quantum field theories, including the 
foundations of the standard
model of particle physics, which have been tested extremely well
up to now, leave little doubt on the validity of 
Special Relativity as a quantum theory of flat space time, 
that is in the absence of gravitational effects. 

One would hope that, like any other successful physical
theory, a physically relevant quantum theory of gravity 
should lead  
to experimental 
predictions that should be testable in the foreseeable future. 
In the QG case, however, 
such predictions may be subtle, due to the extremely weak nature of the 
gravitational interaction as compared with the 
rest of the known interactions in nature. 
Indeed, any hope for 
a true ``phenomenology of QG'' may be easily proven to be
wishful thinking, if all the symmetries and properties that
characterize the current version of what is called particle physics 
phenomenology are still valid in a full quantum theory of gravity.
This is mainly due to the fact that the dimension-full coupling
constant of gravity, the Newton constant $G_N=1/M_P^2$, appears as a  
very strong suppression factor of any physical observable
that could be associated with predictions of quantum gravity,
in a theory in which Lorentz invariance and the laws of quantum mechanics
(or better quantum field theory), such as unitary temporal evolution
and locality of interactions, still hold. 

In recent years, however, more and more physicists contemplate the idea
that such laws, which are characteristic of a flat space time 
quantum field theory, {\emph may not} be valid in a full theory
of curved space times.  
For instance, 
the CPT theorem, one of the most profound results of quantum field 
theory~\cite{cpt}, which is a consequence of Lorentz invariance, 
locality, as well as quantum mechanics (specifically unitary evolution
of a system), may not characterize a quantum gravity theory. 
The possibility of a violation of CPT invariance by quantum gravity 
has been raised in a 
number of theoretical models of QG
that go beyond conventional local 
quantum 
field theoretic treatments of gravity~\cite{ehns,lopez,kostel,bl}.

In a phenomenological, rather model-independent setting, 
CPT Violation has been invoked recently~\cite{mura} in 
an attempt 
to explain
within a three generation scenario, without the introduction of
sterile neutrinos, 
a puzzling experimental result in neutrino physics, namely the 
claims from LSND Collaboration~\cite{LSND} on evidence of 
oscillations in the antineutrino sector 
${\overline \nu}_\mu \to 
{\overline \nu}_e$, due to observed excess of ${\overline \nu}_e$
events, but lack of evidence (depending though on interpretation)
for oscillations in the corresponding 
neutrino sector $\nu_\mu \to \nu_e$.  
In order to explain these results within
the conventional oscillation scenario,
one should invoke different neutrino  mass differences  
in the particle and antiparticle sectors of the model. 
This would
signify CPT violation. Specifically, from solar and atmospheric neutrino data,
the mass squared differences that account for the observed oscillations 
are of order $\Delta m^2 \sim 10^{-5}$ eV$^2$ and $ 10^{-3}$ eV$^2$,
respectively, while the LSND result
would require $\Delta m^2 \sim 10^{-1}$ eV$^2$, which would explain
the suppressed signal in the neutrino sector and the strong signal 
in the antineutrino one. To reconcile, therefore, the LSND result 
with the rest of the available neutrino data, in conventional
oscillation scenaria, one would have to invoke 
CPT violating mass differences between neutrinos and antineutrinos
of order $|m_{\bar{\nu}}^2 - m_{\nu}^2| \sim 10^{-1}$~eV$^2$. 
Although such a CPT violation may look rather drastic, the most stringent
limit we have on the CPT symmetry today, the one coming from the
neutral kaon system, if reinterpreted as a limit on the possible 
differences in mass squared, gives: $|m^2 (K^0) - m^2 (\bar{K}^0)|
< 0.25 $~eV$^2$ showing that such differences are just barely explored
even in the kaon system.
 
Subsequent global analyses of neutrino data~\cite{strumia}, however,  
seem to disfavor two and three generation CPT violating 
scenaria~\cite{mura,bl} thereby leaving  the 
CPT violating scenaria involving sterile neutrinos as the 
only surviving possibility for the explanation of the LSND results. 
Specifically, it has been argued in \cite{barger}, that 
allowing for CPT violation, the mixing matrix elements 
with the forth generation between neutrino
and antineutrino sectors need no longer be the same, 
thereby evading stringent
experimental constraints that killed three-generation models.

Is, however, CPT violation evidenced only as 
an inequality between neutrino-antineutrino masses the only 
way a violation of this symmetry can 
manifest itself in nature? Such a question 
becomes extremely relevant for the case of LSND, because it is
possible that other mechanisms leading to CPT violation exist,
unrelated, in principle, to mass differences between 
particles and antiparticles. Such additional mechanisms for 
CPT violation may well be capable of explaining the LSND results
within a three generation scenario without invoking a sterile neutrino
(a scenario which, on the other hand, is getting totally excluded as
new experimental data become available). 
It is therefore necessary to explore whether alternative ways
exist to account for the LSND result without invoking extra (sterile)
neutrino states.

As we shall argue in this article, quantum 
decoherence may be the key to 
answer this question. Indeed, quantum decoherence in matter propagation
occurs when the matter subsystem interacts with an `environment',
according to the rules of open-system quantum mechanics. 
At a fundamental level,
such a decoherence may be the result of 
propagation of matter in quantum gravity space-time backgrounds
with `fuzzy' properties, which may be responsible for violation
of CPT in a way not necessarily related to mass differences
between particles and antiparticles.

It is the point of this work to claim that decoherence scenaria 
combined with conventional
oscillations,  led as usual by the mass differences between the
mass eigenstates,
can account for all the available neutrino data, including the LSND
result,  within a  three generation model. 
The LSND result would then evidence CPT violation in the sense of  
different decohering interactions between particle and antiparticle
sectors, while the mass differences between the two sectors remain the same.
For detailed reviews of decoherence in two-level systems, with emphasis
on phenomenology including neutrinos,
we refer the reader 
to the literature~\cite{mlambda}. In what follows we shall
only expose the very basic features, which we shall make use of in 
this work in order to perform our phenomenological analysis. 
It is important to stress that in decoherence scenaria 
CPT symmetry is violated in its strong form, in the sense of 
the CPT operator not being well defined. This is an 
important issue which we now come to discuss briefly. 

\section{Quantum Gravity and Decoherence: Brief Review}

A characteristic example of such a violation occurs in 
quantum gravity 
models that 
involve singular space-time 
configurations, integrated over in a path integral formalism,
which are such that the axioms
of quantum field theory, as well as conventional quantum mechanical
behavior, cannot be maintained~\cite{hawking}. Such configurations
consist of wormholes, microscopic (Planck size) black holes, 
and other topologically non-trivial solitonic objects, such as 
{\it geons}~\cite{geons}, {\it etc.} Collectively, we may call 
such configurations {\it space time foam}, a terminology given by 
 J.A. Wheeler~\cite{wheeler} who first conceived 
the idea that 
the structure of the {\it quantum} space time at Planck scales ($10^{-35}$ m)
may indeed be fuzzy.

It has been argued that, as result, 
a mixed state description must be 
used ({\it QG-induced decoherence})~\cite{hawking,ehns}, 
given that such objects
cannot be accessible to low-energy observers, and as such must be traced over
in an effective field theory context. For the case of microscopic black holes
this can be readily understood by the loss of information across microscopic
event horizons, as a corollary of which it has been argued~\cite{wald},
already back in 1978, 
that CPT invariance in its strong form  
must be abandoned in a foamy quantum gravity theory.
Such a breakdown of CPT symmetry is a fundamental 
one, and, in particular, implies that a proper CPT 
operator may be {\it ill defined} in such QG decoherence cases. 

The main reason behind such a failure is the non factorisability of the 
super-scattering matrix, \$, connecting asymptotic ``in'' and ``out''
density matrices~\cite{hawking}. The latter describe 
{\it mixed} quantum states due to  
the existence of a quantum-gravitational ``environment'' of microscopic
singular space-time configurations. The presence of the environment
leads to {\it decoherence}, like in any other case of 
an open quantum-mechanical system,
implying an evolution from an initial pure quantum mechanical state $|\phi>$ 
to a mixed state $\rho_\psi = {\rm Tr}|\psi><\psi|$ (the trace is over
unobserved degrees of freedom, `lost' in the environment). 
This in turn implies~\cite{wald} the
{\it non invertibility} of the \$-matrix, as a quantum mechanical operator
acting on density matrices. 
The associated CPT non invariance is then linked to the 
fact that if CPT were a well defined operator, stemming from the invariance
of the system under its action, then \$ would have been invertible, in 
obvious contradiction with the information loss and decoherence
inherent to the problem. 

As emphasized recently in \cite{bmp}
this would also imply that the concept of {\it antiparticle}, 
may be {\it ill defined}, with interesting 
testable consequences in multiparticle situations, 
such as 
those encountered 
in (EPR-like entangled) meson factories. Such CPT violation
is distinct from other violations of this symmetry as a result
of Lorentz symmetry breakdown
in the so-called Standard Model extension (SME)~\cite{kostel}, 
or locality violations~\cite{bl}, which have 
been considered in the recent literature as other 
experimentally testable approaches to quantum gravity. 
It should be noted that 
these latter violations of symmetries may or may not coexist 
with QG decoherence, given that Lorentz invariance is 
not necessarily inconsistent with decoherence~\cite{millburn}.  

Some caution should be paid regarding CPT Violation through decoherence.
As emphasized in \cite{wald}, from a formal view point, 
the non-invertibility of the 
\$-matrix, which 
implies a strong violation of CPT in the sense described above,
does not preclude a softer form of CPT invariance, 
in the sense that any strong form of CPT violation does not 
necessarily have to show 
up in any single  experimental measurement. This implies that, despite
the general evolution of pure to mixed states, it may still be possible
in the laboratory to ensure that the system evolves 
from an initial pure state $\psi$ to a single final 
state $\phi$, and that the weak form of CPT invariance is manifested 
through the equality of probabilities between the states $\psi$, $\phi$:
\begin{equation} 
P(\psi \to \phi) = P(\theta^{-1}\phi \to \theta \psi)~,
\label{probs}
\end{equation} 
with $\theta$ a CPT operator acting on the subspace of pure state vectors
in the laboratory, such that for density matrices 
$\Theta \rho = \theta \rho \theta^\dagger$,
$\theta^\dagger = -\theta^{-1}$, and \$$^\dagger = \Theta^{-1}$\$$\Theta^{-1}$,
but \$ $^\dagger \ne $\$ $^{-1}$, unless full CPT invariance holds, which is 
the case of 
no decoherence. If this is the case, then the decoherence-induced
CPT violation will not show up in any experimental measurement.
It is therefore important to check whether it is possible to 
test experimentally which case is realized in nature. 
Fortunately this can be done in a clear way. 

Indeed, the main consequence of the quantum-gravity induced  
decoherence is the {\it modified}
non-quantum mechanical evolution of a matter system under
consideration, as a result of the interaction 
with the QG environment~\cite{ehns}.
Such an evolution will result in many observable consequences,
which may be testable in the near future with a precision 
that in some cases can reach Planck scale sensitivity. 
Historically, the first probe of such non-quantum mechanical
decoherence and  
CPT violating behavior (in the sense of the strong form violation 
of \cite{wald}) is the neutral kaon system~\cite{ehns,lopez},
in both single-particle experiments, such as the CPLEAR~\cite{cplear}
and KTeV experiments~\cite{ktev}, as well as 
multi-particle situations, such as kaon ($\phi$) factories~\cite{dafne}.
Other neutral mesons, such as B mesons~\cite{bmeson}, 
can also be used as sensitive probes of QG decoherence. 

In single-particle neutral meson experiments 
one looks at time profiles of asymmetries~\cite{lopez}.
The mass difference between the energy eigenstates $K_L, K_S$  
is responsible for the `bulk' behavior of the asymmetry, 
for instance leading to an
oscillatory pattern in some of the asymmetries, such as $A_{2\pi}$, 
decoherence on the other hand 
can be  responsible for a {\it slight distortion} of such a 
behavior~\cite{lopez}. The non observation of such distortions
results in stringent bounds on the decoherence parameters. 
Mathematically, one does not have to have a detailed knowledge
of the dynamics of quantum gravity in order to perform experimental
low-energy tests of decoherence. This can be achieved following the 
so-called Lindblad or mathematical semi-groups approach
to decoherence~\cite{lindblad},
which is a very efficient way of studying open systems
in quantum mechanics. The time irreversibility in the evolution
of such semigroups, which is linked to decoherence, 
is inherent in the mathematical property 
of the lack of an inverse in the semigroup. This approach has been 
followed for the study of quantum-gravity decoherence in the 
case of neutral kaons in \cite{ehns,lopez}. 

In the parameterization of 
\cite{ehns} for the decoherence effects, 
using three 
decoherence parameters with dimensions of energy, 
$\alpha,\beta,\gamma$, one finds the following 
bounds from the CPLEAR experiment~\cite{cplear}: 
\begin{eqnarray}
&& \alpha < 4 \times 10^{-17}~{\rm GeV}, |\beta| < 2.3 \times 10^{-19}~{\rm GeV}, \nonumber \\
&&\gamma < 3.7 \times 10^{-21}~{\rm GeV}
\end{eqnarray} 
where the positivity of $\rho$,
required by the fact that its diagonal elements 
express probability densities, implies $\alpha, \gamma \ge 0$, and 
$\alpha \gamma \ge \beta^2$.

The Lindblad approach to decoherence does not require any detailed knowledge
of the environment, apart from energy conservation, 
entropy increase and complete positivity of the (reduced) density 
matrix $\rho(t)$ of the subsystem under consideration. 
The basic evolution equation for the (reduced) density matrix of the 
subsystem in the Lindblad approach is {\it linear} in $\rho(t)$ and 
reads: 
\begin{equation} 
\frac{\partial \rho}{\partial t} = -i[H_{\rm eff}, \rho] + \frac{1}{2}\sum_{j}\left([b_j, \rho(t)b^\dagger_j] + [b_j\rho(t), b_j^\dagger]\right)~,
\label{lindblad}
\end{equation}
where $H_{\rm eff}$ is the effective Hamiltonian of the subsystem, 
and the operators $b_j$ represent the interaction with the environment,
and are assumed bounded. Notice that the Lindblad part
cannot be written as a commutator (of a Hamiltonian 
function) with $\rho$. Environmental 
contributions that 
can be cast in Hamiltonian evolution (commutator form) are absorbed
in $H_{\rm eff}$.

It must be noted at this stage that the requirement
of complete positivity, which essentially
pertains to the positivity of the map $\rho(t)$ as the time evolves
in the case of many particle situations, such as meson factories 
(two-kaon states ($\phi$-factory), or two-$B$-meson states {\it etc.}),
may not be an exact property of quantum gravity, whose interactions
with the environment could be {\it non linear}~\cite{emn}. 
Nevertheless, complete 
positivity leads to a convenient and simple parametrization, and 
it has been assumed so far in many phenomenological 
analyses of quantum gravity decoherence in generic two state systems,
such as two-flavor neutrino systems~\cite{lisi,benatti2}.
In fact, in the parametrization of \cite{ehns}, 
the imposition of complete positivity 
leads to $\alpha=\beta=0$, leaving only $\gamma >0$ as the 
only decoherence parameter in the two-state system~\cite{benatti}.

Formally, the bounded Lindblad operators of an $N$-level quantum 
mechanical system can be expanded in a basis of matrices 
satisfying standard commutation relations of Lie groups. 
For a two-level system~\cite{ehns,lopez} such matrices are the 
SU(2) generators (Pauli matrices) 
plus the $2\times 2$ identity operator, while for a three level
system~\cite{gago}, which will be relevant for our purposes in this article,
the basis comprises of  
the eight Gell-Mann $SU(3)$ matrices $\Lambda_i~,~i=1,\dots 8$
 plus the $3\times 3$ identity matrix $I_{3x3}$. 

Let ${\cal J}_\mu$, $\mu=0,\dots 8 (3)$ be a set of SU(3) (SU(2)) generators  
for a three(two)-level system; then, one may expand the various terms 
in (\ref{lindblad}) in terms of ${\cal J}_\mu $ to arrive at the generic form:
\begin{eqnarray}
&&\frac{\partial \rho_\mu}{\partial t} = \sum_{ij} h_i\rho_j {f}_{ij\mu}  
+ \sum_{\nu} {\cal L}_{\mu\nu}\rho_\mu~, \nonumber \\
&& \mu, \nu = 0, \dots N^2 -1, \quad i,j = 1, \dots N^2 -1
\label{expandedlind}
\end{eqnarray}
with $N=3 (2)$ for three(two) level systems, and $f_{ijk}$ the structure
constants of the $SU(N)$ group. 
The requirement for entropy increase implies the hermiticity of  the 
Lindblad operators $b_i$, as well as the fact that 
the matrix ${\cal L}$ of the the non-Hamiltonian part of the 
evolution has the properties that ${\cal L}_{0\mu}={\cal L}_{\mu 0} =0$,
${\cal L}_{ij} =\frac{1}{2}\sum_{k,\ell ,m} b_m^{(n)} b_k^{(n)}{f}_{imk}{f}_{\ell k j}$, with the notation $b_j \equiv \sum_\mu b_\mu^{(j)} {\cal J}_\mu $.   

In the two-level case of \cite{ehns} the decoherence matrix ${\cal L}_{\mu\nu}$ 
is parametrized by a $4\times 4$ matrix, whose non vanishing entries 
are occupied by the three parameters with the dimensions of energy 
$\alpha, \beta, \gamma$ with the properties mentioned above. 
If the requirement of a completely positive map $\rho (t)$ is imposed, 
then the $4 \times 4$ matrix ${\cal L}$ 
becomes diagonal, with only one non vanishing 
entry occupied by the decoherence parameter $\gamma > 0$~\cite{benatti}. 

\section{Two Generation Neutrino Models and Decoherence: Brief Review}

If the above formalism is applied to two-level neutrino oscillation physics,
in an attempt to discuss the influence of a quantum-gravity environment on 
neutrino oscillations, then there are two possible forms of the matrix
${\cal L}$: ${\cal L}={\rm Diag}\left(0, 0, -\gamma, 0\right)$ in the case
where energy and lepton number are {\it conserved},  and 
${\cal L}={\rm Diag}\left(0, 0, 0, -\gamma\right)$ in the case that 
energy and lepton number  are violated by quantum gravity, 
but {\it flavor} is conserved. In \cite{lisi} the energy conserving 
case has been considered, with the conclusion that pure decoherence,
i.e. induced oscillations in the absence of neutrino mass differences, 
was incompatible 
with the observed phenomenology of neutrino oscillations. 

This is in agreement with the kaon case~\cite{lopez}, where again 
the observed bulk behavior of the time profiles of the various 
kaon asymmetries, as well as the observed CP violation,
cannot be explained solely by quantum decoherence, the latter
providing only distortions to the `bulk' behavior.
Including the mass-induced oscillations, combined with decoherence, 
and 
comparing
with atmospheric neutrino data, 
the authors of \cite{lisi} managed to obtain bounds for  
the decoherence parameter $\gamma$ 
in the presence of mass differences (oscillations); 
as in the kaon 
case~\cite{lopez}, such bounds are obtained merely 
by the non-observation 
of the 
slight distortions to 
the
mass-induced neutrino oscillations 
that the presence of decoherence 
would cause.

The so-obtained limits for the decoherence Lindblad parameter $\gamma$  
in \cite{lisi} can be summarized as follows:
adopting the parametrization $\gamma = \gamma_0 (E/{\rm GeV})^n$, 
they considered three cases: (a) n=0, in which the experimental
data imply $\gamma_0 < 10^{-23}$~GeV, (b) 
n=2, which is the case expected in some
rather optimistic (from the prospect of experimental
detection) theoretical models 
of quantum gravity~\cite{emn,lopez},  for which 
the most stringent bound  derived  is  $\gamma_0 < 0.9 \times 10^{-27}$,
and (c) n=-1, which is the case in which the decoherence parameter
exhibits an  
energy dependence that mimics the conventional oscillation scenario, 
for which $\gamma_0 < 2 \times 10^{-21}$~GeV$^2$. 

It is worth  noticing, that as far as  case (b) is concerned, at least
by dimensional considerations, one is tempted to assume theoretical
values of the decoherence parameter which are of the form $E^2/M_{QG}$,
with $M_{QG}$ the effective quantum gravity scale `felt' by the neutrinos.
The above bounds indicate a value $M_{QG} \sim 10^{27}$ GeV
for $E ={\cal O}({\rm few~GeV})$ which  
are typical in atmospheric neutrino experiments. This is much higher  
than
the theoretically expected (on naturalness grounds) Planck value 
$M_{QG} \sim 10^{19}$~GeV. This is to be contrasted with the 
case of kaons~\cite{ehns}, where, as mentioned above,  
the bound on the decoherence parameter $\gamma < 10^{-21}$GeV
obtained from CPLEAR~\cite{cplear},
was found to be much closer to 
the Planck scale, the natural QG scale.  
On the other hand, the case (c), with an inverse energy $E$ dependence,
is purely phenomenological and does not stem directly from a 
specific model of decoherence. The bound on $\gamma_0 < 10^{-23}$~(GeV)$^2$ 
obtained in that case is of the order of the quantity $\Delta m^2 /E$
for atmospheric neutrino energies, and
square mass differences of order $10^{-5}$~eV$^2$ which is assumed 
to be the right order for solar neutrino oscillations in the 
conventional oscillation scenario. 

Unfortunately, the analysis of the data presented in \cite{lisi}
within a two-generation scenario 
did not show any concrete evidence for such a type-(c) decoherence, 
even after inclusion~\cite{lisi2} of the 
first K2K data~\cite{K2K}, in combination 
with Super Kamiokande (SK) data~\cite{SK}. 
This prompted these authors to draw the pessimistic conclusion
that the extension of the 
neutrino oscillation plus decoherence 
scenario to the fully-fledged case of
three neutrino generations was not worthy. 
Adding to this pessimism, 
there were also some theoretical estimates of the 
decoherence parameter in {\it specific} models of quantum-gravity 
induced decoherence for generic two-level systems~\cite{adler}, 
according to which 
$\gamma_0 \sim (\Delta m^2 )^2/E^2M_P $, with $\Delta m^2$ the mass-squared 
difference between, say, the two neutrinos (or kaons etc). 
For $M_P \sim 10^{19}$~GeV,
the Planck scale, this is beyond any prospects for experimental
detection in oscillation experiments in the foreseeable future.
However, as remarked in \cite{emn}, such estimates are specific 
to the model considered in \cite{adler}, and indeed there are concrete
examples of theoretical (stringy) models of 
quantum-gravity-induced decoherence where they are not applicable.
We shall discuss such issues later on in our article.

\section{Extension to three generations: Combining Oscillations with 
Decoherence}

Pessimistic thoughts have also appeared in the work of 
\cite{gago}, which considered the extension of the completely
positive decoherence scenario to the standard three-generation neutrino 
oscillations case. 
The above-described three-state Lindblad problem has been adopted
for the discussion of this case. The relativistic neutrino Hamiltonian
$H_{\rm eff} \sim p^2 + m^2/2p$, with $m$ the neutrino mass,  
has been used as the Hamiltonian
of the subsystem in the evolution of eq.(\ref{lindblad}).

In terms of the generators ${\cal J}_\mu$, $\mu = 0, \dots 8$ 
of the SU(3) group, $H_{\rm eff}$ can be expanded as~\cite{gago}:
${\cal H}_{\rm eff} = \frac{1}{2p}\sqrt{2/3}\left(6p^2 + \sum_{i=1}^{3}m_i^2
\right){\cal J}_0 + \frac{1}{2p}(\Delta m_{12}^2){\cal J}_3
+ \frac{1}{2\sqrt{3} p}\left(\Delta m_{13}^2 + \Delta m_{23}^2 \right){\cal J}_8$, 
with the obvious notation $\Delta m_{ij}^2 = m_i^2 - m_j^2$, $i,j =1,2,3$.

The analysis of \cite{gago} assumed {\it ad hoc} a diagonal form 
for the $9 \times 9$ decoherence matrix ${\cal L}$ in (\ref{expandedlind}):
$[{\cal L}_{\mu\nu}]= {\rm Diag}\left(0, -\gamma_1,-\gamma_2,-\gamma_3,-\gamma_4,-\gamma_5,-\gamma_6,-\gamma_7,-\gamma_8\right)$ in direct analogy with 
the two-level case of complete positivity~\cite{lisi,benatti}. 
As we have mentioned already, there is no strong physical
motivation behind such restricted forms of decoherence. This assumption,
however,
leads to  the simplest possible decoherence models, 
and, for our  
{\it phenomenological} purposes
in this work,  
we will assume 
the above form, which we will use
to fit all the available neutrino data. It must be clear to the reader though,
that such a simplification, if proven to be successful (which, as we shall 
argue below, is the case here),
just adds more in favor of decoherence models, 
given the 
restricted number of available parameters for the fit in this case. 
In fact, any other non-minimal
scenario will have it easier to accommodate data 
because it will have more degrees of freedom available for such a 
purpose.

Specifically we shall look at transition probabilities~\cite{gago}:
\begin{eqnarray} 
&& P(\nu_\alpha \to \nu_\beta) ={\rm Tr}[\rho^\alpha (t)\rho^\beta] 
= \nonumber \\
&&\frac{1}{3} + \frac{1}{2}\sum_{i,k,j}
e^{\lambda_k t}{\cal D}_{ik}{\cal D}_{kj}^{-1}\rho^\alpha_j (0)\rho_i^\beta
\label{trans}
\end{eqnarray}
where $\alpha,\beta = e, \mu, \tau$ stand for the three neutrino flavors, and 
Latin indices run over $1, \dots 8$. The quantities $\lambda_k$ 
are the eigenvalues of the matrix ${\cal M}$ appearing in the 
evolution (\ref{expandedlind}), after taking into account probability
conservation, which decouples $\rho_0(t)=\sqrt{2/3}$, leaving the remaining 
equations in the form: $\partial \rho_k /\partial t = \sum_{j} 
{\cal M}_{kj}\rho_j $. The matrices ${\cal D}_{ij}$ are the matrices
that diagonalize ${\cal M}$~\cite{lindblad}. Explicit forms of 
these matrices, the eigenvalues $\lambda_k$, 
and consequently the transition probabilities
(\ref{trans}), are given in \cite{gago}. 

The important point to stress is that, in generic models of oscillation plus 
decoherence, the eigenvalues $\lambda_k$ depend
on both the decoherence parameters $\gamma_i$ and the mass differences
$\Delta m^2_{ij}$. For instance, $\lambda_1 = \frac{1}{2}[-(\gamma_1 + \gamma_2)
-\sqrt{(\gamma_2 - \gamma_1)^2 -4\Delta_{12}^2}]$, with 
the notation $\Delta_{ij} \equiv \Delta m_{ij}^2/2p$, $i,j=1,2,3$.
 Note that, to leading order in the (small) squared-mass differences, 
one may replace
$p$ by the total neutrino energy $E$, and this will be understood 
in what follows.
For detailed expressions of the rest of the parameters 
we refer the reader to \cite{gago}. However, we note, for future use, that 
it is a generic feature of the $\lambda_k$ to depend on  
the quantities 
$\Omega_{ij}$ which are given by
\begin{eqnarray}
\Omega_{12} &=& \sqrt{(\gamma_2 - \gamma_1)^2 -4\Delta_{12}^2} \nonumber \\
\Omega_{13} &=& \sqrt{(\gamma_5 - \gamma_4)^2 -4\Delta_{13}^2} \\
\Omega_{23} &=& \sqrt{(\gamma_7 - \gamma_6)^2 -4\Delta_{23}^2} \nonumber 
\end{eqnarray}
{}From the above expressions  
for the eigenvalues $\lambda_k $, it becomes clear
that, when decoherence and oscillations
are present simultaneously,
one should distinguish two cases, 
according to the relative magnitudes of $\Delta_{ij}$ and 
$\Delta\gamma_{kl} \equiv \gamma_k - \gamma_l$: (i) $2|\Delta_{ij}| \ge |\Delta\gamma_{k\ell}|$, and (ii) 
$2|\Delta_{ij}| < |\Delta\gamma_{k\ell}|$.
In the former case, the probabilities (\ref{trans}) contain trigonometric
(sine and cosine) functions, whilst in the latter they exhibit  
hyperbolic sin and cosine dependence.

Assuming mixing between the flavors, amounts to expressing neutrino
flavor eigenstates $|\nu_\alpha>$, $\alpha=e,\mu,\tau$ in terms of 
mass eigenstates $|\nu_i>$, $i=1,2,3$ through a (unitary) matrix $U$:
$|\nu_\alpha> = \sum_{i=1}^{3}U^*_{\alpha i}|\nu_i>$. This implies that
the density matrix of a flavor state $\rho^\alpha $ can be expressed 
in terms of mass eigenstates as: 
$\rho^\alpha=|\nu_\alpha><\nu_\alpha|=\sum_{i,j}U^*_{\alpha i}U_{\alpha j}|\nu_i><\nu_j| $. From this we can determine 
$\rho_\mu^\alpha =2{\rm Tr}(\rho^\alpha {\cal J}_\mu )$, a quantity needed to
calculate the transition probabilities (\ref{trans}).

The important comment 
we would like to raise at this point is that, when considering
the above probabilities in the antineutrino
sector, the respective decoherence parameters ${\bar \gamma}_i$ 
in general may be different from the corresponding ones in the 
neutrino sector,
as a result of the strong form of CPT violation. In fact, as we 
shall discuss next, this will be crucial for accommodating 
the LSND result without conflicting with  the rest of the available neutrino data.
This feature is totally unrelated to mass differences between flavors. 

\section{Combining Decoherence, Oscillations and the LSND Result}

In \cite{gago} a pessimistic conclusion was drawn on 
the ``clear incompatibility between neutrino data and theoretical
expectations", as followed by their qualitative tests for decoherence.
It is a key feature of our present work 
to point out that we do not share at all this point of view. 
In fact, as we shall demonstrate below, 
if one takes into account  all the available 
neutrino data, including the final 
LSND results~\cite{LSND}, which the authors of \cite{gago} did not do,
and allows for the above mentioned CPT violation in the decoherence
sector, then one will arrive at exactly the opposite conclusion, namely 
that three-generation decoherence and oscillations can fit 
the data successfully!
 
As we shall argue here, compatibility of all available 
data, including CHOOZ~\cite{chooz} and LSND, can be achieved through a  
set of decoherence parameters $\gamma_j$ with energy dependences 
$\gamma^0_j E$  and $\gamma^0_j/E$,  
with $\gamma^0_j  
\sim 10^{-18}, 10^{-24}$~(GeV)$^2$, respectively, 
for some $j$'s. 
As it will be clear later, the quality of the fit
diminishes to the level of the standard three-oscillation case once
KamLAND spectral distortion data~\cite{kamspec} 
are included but is significanlty higher
otherwise, showing that either additional energy dependences are
needed or non-diagonal elements are required to get better agreement with data.

Some important remarks are in order. First of all, 
in the analysis of \cite{gago} 
pure decoherence is excluded
in three-generation scenaria, as in two generation ones, 
due to 
the fact that the transition probabilities in the case $\Delta m_{ij}^2=0$
(pure decoherence) are such that the survival probabilities 
in both sectors are equal, i.e. 
$P(\nu_\alpha \to \nu_\alpha ) =
P({\overline \nu}_\alpha \to {\overline \nu}_\alpha) $. 
{}From (\ref{trans}) we have in this case~\cite{gago}:
\begin{eqnarray} \label{trans2}
P_{\nu_e \to \nu_e} = P_{\nu_\mu \to \nu_\mu}
\simeq \frac{1}{3} + \frac{1}{2}e^{-\gamma_3 t}
+ \frac{1}{6}e^{-\gamma_8 t} 
\end{eqnarray}
{}From the CHOOZ experiment~\cite{chooz}, for which $L/E \sim 10^3/3$~m/MeV,
we have that $\langle P_{{\bar \nu}_e \to {\bar \nu}_e} \rangle \simeq 1$, while the K2K 
experiment~\cite{K2K} with $L/E \sim 250/1.3 $~ km/GeV has observed events
compatible with $\langle P_{\nu_\mu \to \nu_\mu} \rangle \simeq 0.7$,
thereby contradicting the theoretical predictions 
(\ref{trans2}) of pure decoherence. 

However, this conclusion is based on the fact that in the antineutrino sector
the decoherence matrix is the same as that in the neutrino sector. 
In general this need not be the case, in view of CPT violation, which 
could imply a different interaction 
of the antiparticle with the gravitational
environment as compared with the particle.

In our tests we take into account this possibility, but as can be seen
from the figure,  and will be evident later on, once the analysis
will be explained in some depth,
pure decoherence can be excluded also in this case, 
as it is clearly  incompatible
with the totality of the available data.

In order to check our model, we have performed a 
$\chi^2 $ comparison  (as opposed to a $\chi^2 $ fit)
to SuperKamiokande sub-GeV and multi GeV data
(the forty data points that are shown in the figures), CHOOZ
data (15 data points) LSND (1 datum) and KamLAND spectral 
information~\cite{kamspec}
(sampled in 13 bins), for a sample point in the
vast parameter space of our 
extremely simplified version of decoherence models. 
Let us emphasize 
that we have {\bf not} performed a $\chi^2$-fit and therefore the point
we are selecting (by ``eye'' and not by $\chi$) is not optimized to give
the best fit to the existing data. Instead, it must be regarded 
as one among the many equally good sons in this family of solutions,
being extremely possible to find a better fitting one through a 
complete (and highly time consuming) scan over the whole parameter 
space.  

Cutting the long story short, and to make the analysis easier, we have set
all the $\gamma_i$ in the neutrino sector to zero, restricting this way,
all the decoherence effects to the antineutrino one where, we have assumed
for the sake of simplicity,
\begin{eqnarray} 
&&\bar{\gamma}_i = \bar{\gamma}_{i+1}~{\rm for}~~ i=1, 4, 6, 7~
{\rm and}~ \bar{\gamma}_1 = \bar{\gamma}_4~,
\bar{\gamma}_3 = \bar{\gamma}_{8} 
\label{decpars}
\end{eqnarray} 
Furthermore, we have also set the 
CP violating phase of the NMS matrix to zero, so that all the mixing
matrix elements become real.

With these assumptions, the otherwise cumbersome 
expression for the transition
probability for the antineutrino sector takes the form,
\begin{eqnarray}
  P_{\bar\nu_{\alpha}\rightarrow \bar\nu_{\beta}} & = &\frac{1}{3}+\frac{1}{2}\left\{ 
\rho_{1}^{\alpha}\rho_{1}^{\beta}
\cos\left(\frac{|\Omega_{12}| t}{2}\right)
e^{-\bar\gamma_{1} t} \right. \nonumber \\
&+ &  \rho_{4}^{\alpha}\rho_{4}^{\beta}
\cos\left(\frac{|\Omega_{13}|t}{2}\right)  e^{-\bar\gamma_{4} t} \nonumber \\
&+& \rho_{6}^{\alpha}\rho_{6}^{\beta} 
\cos\left(\frac{|\Omega_{23}|t}{2}\right) e^{-\bar\gamma_{6} t} 
\nonumber \\
&+ &   e^{-\bar\gamma_{3}t} \left( \rho_{3}^{\alpha}\rho_{3}^{\beta}+
\rho_{8}^{\alpha}\rho_{8}^{\beta} \right) \Bigg\}. 
\label{prob}
\end{eqnarray}
where the $\Omega_{ij}$ were 
defined in the previous section 
and are the same in both sectors (due to our choice of $\gamma_i$'s) and
\begin{eqnarray}\label{mixpar}
& & \rho_{1}^{\alpha}=2\;\re(U_{\alpha 1}^{\ast}U_{\alpha 2}) \nonumber \\
& & \rho_{3}^{\alpha}=|U_{\alpha 1}|^{2}-|U_{\alpha 2}|^{2} \nonumber \\
& & \rho_{4}^{\alpha}=2\;\re(U_{\alpha 1}^{\ast}U_{\alpha 3}) \\
& & \rho_{6}^{\alpha}=2\;\re(U_{\alpha 2}^{\ast}U_{\alpha 3}) \nonumber \\
& & \rho_{8}^{\alpha}=\frac{1}{\sqrt{3}}(|U_{\alpha 1}|^{2}+|U_{\alpha 2}|^{2}-2|U_{\alpha 3}|^{2}) \nonumber
\end{eqnarray}
where the mixing matrices are the same as in the 
neutrino sector.
For the neutrino sector, as there are no decoherence effects, the standard
expression for the transition probability is valid.

It is obvious now that, since the neutrino sector does not suffer from
decoherence, there is no need to include the solar data into the fit.
We are guaranteed to have an excellent agreement with solar data, as long
as we keep the relevant mass difference and mixing angle within
the LMA region, something which we shall certainly do.

As mentioned previously, CPT violation is driven by, and restricted to, the
decoherence parameters, and hence masses and mixing angles are
the same in both sectors, and selected to be 

\centerline{$\Delta m_{12}^2 = \Delta \overline{ m_{12}}^2 = 
7 \cdot 10^{-5}$~eV$^2$,}

\centerline{$\Delta m_{23}^2 = \Delta \overline{ m_{23}}^2 = 
2.5 \cdot 10^{-3}$~eV$^2$,}

\centerline{$\theta_{23} = \overline{\theta_{23}}= \pi/4$, $\theta_{12} = 
\overline{\theta_{12}}= .45$,}

\centerline{$\theta_{13} = \overline{\theta_{13}}= .05$,} 

\noindent 
as indicated by the state of the art analysis.

For the decoherence parameters we have chosen (c.f. (\ref{decpars})) 
\begin{eqnarray} 
 \overline{\gamma_{1}} = \overline{\gamma_{2}} = \overline{\gamma_{4}} &=& 
\overline{\gamma_{5}} =
2 \cdot 10^{-18} \cdot E
\nonumber \\
& {\rm and }& \nonumber \\
 \overline{\gamma_{3}} =  \overline{\gamma_{6}} = \overline{\gamma_{7}} &=& 
\overline{\gamma_8}=
 1 \cdot 10^{-24}/E~, 
\label{decohparam}
\end{eqnarray}
where 
$E$ is  the  neutrino energy, 
and barred quantities
refer to the antineutrinos, given that decoherence takes place only 
in this 
sector in our model.
All the other parameters 
are assumed to be zero.  All in all, we have introduced only
two new parameters, two new degrees of freedom, $ \overline{\gamma_{1}}$
and $ \overline{\gamma_{3}}$,
and we shall try
to explain with them all the available experimental data.

It can be checked straightforwardly, by means of (\ref{prob}),
that for the regime of the parameters relevant to the 
experiments we are considering in this work,
this parametrization guarantees the positivity of the 
relevant probabilities.
This is an important issue, because usually negative probabilities
are viewed as a signal of inconsistency in parameterizing 
the pertinent decoherence effects. In our case, with two different
energy dependences in the (non constant) decoherence coefficients, 
there seems to be no master condition that guarantees
the complete positivity in a way independent 
of the energies/momenta of the neutrinos. As we shall discuss later 
on in the article, we attribute the $1/E$-dependent decoherent 
coefficients to 
conventional matter effects, while the $E$-dependent ones are associated
with novel effects of quantum gravity, which increase with the energy 
of the probe. The negative probabilities that may occur outside the
regime of parameters to be specified below are then interpreted
as implying simply 
that our linear (and simplified) parametrization of 
quantum gravity decoherence (\ref{decohparam}) 
ceases to be valid in such regimes, and more complicated,
probably non linear, entanglement may be in operation,
as expected in a quantum theory of gravity~\cite{emn}.

With these in mind, we now observe that, in our simplified model 
of decoherence, parametrized by a diagonal decoherence
matrix, positivity of the relevant probabilities seems to be guaranteed
for $\overline \gamma_1 L > \overline \gamma_3 L $, which, with our 
parametrization (\ref{decohparam}), implies: $L/E \le 10^{24}-10^{25}$ 
GeV$^{-2}$,
and $E > 1$ MeV. 
These sufficient conditions are met by 
all the  current and planned terrestrial (anti)neutrino experiments.
Outside this regime, our parametrization simply fails, and one needs to 
resort to more complicated situations, which fall beyond the 
scope of the present paper, and will be presented elsewhere.
                         
At this point it is important to stress that the inclusion of two 
new degrees of freedom 
is not sufficient to guarantee that one will indeed be able to
account for all the experimental observations.
We have
to keep in mind that, in no-decoherence
situations, the addition of a sterile neutrino (which comes
along with four new degrees of freedom -excluding again the possibility
of CP violating phases) did not seem to be sufficient for 
matching 
all
the available experimental data, at least in CPT conserving situations.  

In order to test our model 
with these two decoherence parameters in the antineutrino sector,  
we have calculated the zenith angle dependence of the 
ratio ``observed/(expected in the no oscillation case)'', for muon and
electron atmospheric neutrinos, for the sub-GeV and multi-GeV energy ranges,
when mixing is taken into account. Since matter effects are important for
atmospheric neutrinos, we have implemented them through a 
two-shell model, where 
the density in the mantle (core) is taken to be roughly 3.35 (8.44) gr/cm$^3$,
and the core radius is taken to be 2887 km. 
We should note at this stage that a ``fake'' CPT Violation 
appears due to matter
effects, arising from a relative sign difference of the matter
potential between the respective interactions of 
neutrinos and antineutrinos with ordinary matter. This, however, 
is easily disentangled from our genuine (due to quantum gravity)
CPT Violation, used here to parametrize our model fit to LSND results; 
indeed, a systematic study 
of such effects~\cite{ohlsson} has shown that ``fake'' CPT
Violation increases with the oscillation length, but decreases
with the neutrino energy, $E$, vanishing in the limit $E \to \infty$;  
moreover, no independent information
regarding such effects 
can be obtained by looking at the antineutrino sector, as
compared with data from the neutrino sector,  
due to the fact that in the presence of ``fake'' CPT Violation, 
but in the absence of any genuine CPT breaking, 
the pertinent CPT probability differences between 
neutrinos and antineutrinos are related, 
$\Delta P^{\rm CPT}_{\alpha\beta} = 
-\Delta P^{\rm CPT}_{\overline{\beta}\overline{\alpha}}$, 
where $\Delta P_{\alpha\beta}^{\rm CPT}=P_{\alpha\beta}-P_{\overline \beta \overline \alpha}$, and the Greek indices denote neutrino flavors.
These
features are to be contrasted with our dominant decoherence
effects $\overline \gamma_1 $ (\ref{decohparam}), 
proportional to the antineutrino energy, $E$, which are dominant 
only in the antineutrino sector. For the same reason, 
our effects can be disentangled from ``fake'' decoherence 
effects arising from Gaussian averages of the oscillation 
probability due to, say, uncertainties in the energy 
of the neutrino beams~\cite{ohlsson2}, which are the same
for both neutrinos and antineutrinos. We, therefore, claim that the
complex energy dependence in (\ref{decohparam}), 
with {\it both} $L\cdot E$ and $L/E$ terms being present in the 
antineutrino sector, 
may be a characteristic  feature of new physics, with the $L \cdot E$ 
terms being related to quantum-gravity induced 
(genuine) CPT Violating decoherence. 

The results are shown in Fig. 1 (c),  
where, for the sake of comparison,
we have also included 
the experimental data.  
We also present in that figure the pure decoherence scenario in the
antineutrino sector (a), as well as in both sectors (b).  
For completeness, we also present a scenario with neutrino mixing
but with 
decoherence 
operative in both sectors (d).  The conclusion is straightforward:
while pure decoherence appears to be excluded,decoherence plus mixing
provides an astonishing agreement with experiment\footnote{Pure decoherence
models, are not ruled out by the present analysis, which tests only one
particular point in a huge parameter space, but by a complete scan 
like the ones in \cite{gago,lisi2}. }.

\begin{figure}[tb]
\includegraphics[width=9cm]{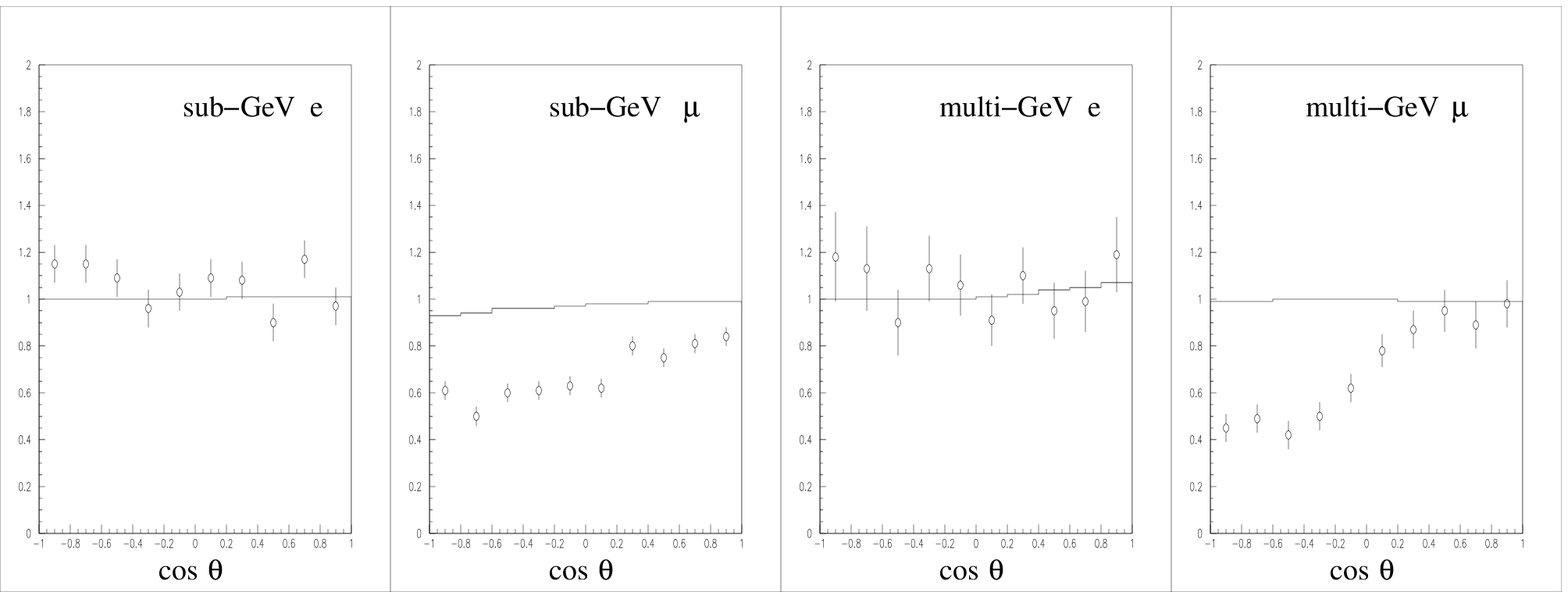} \hfill 
\includegraphics[width=9cm]{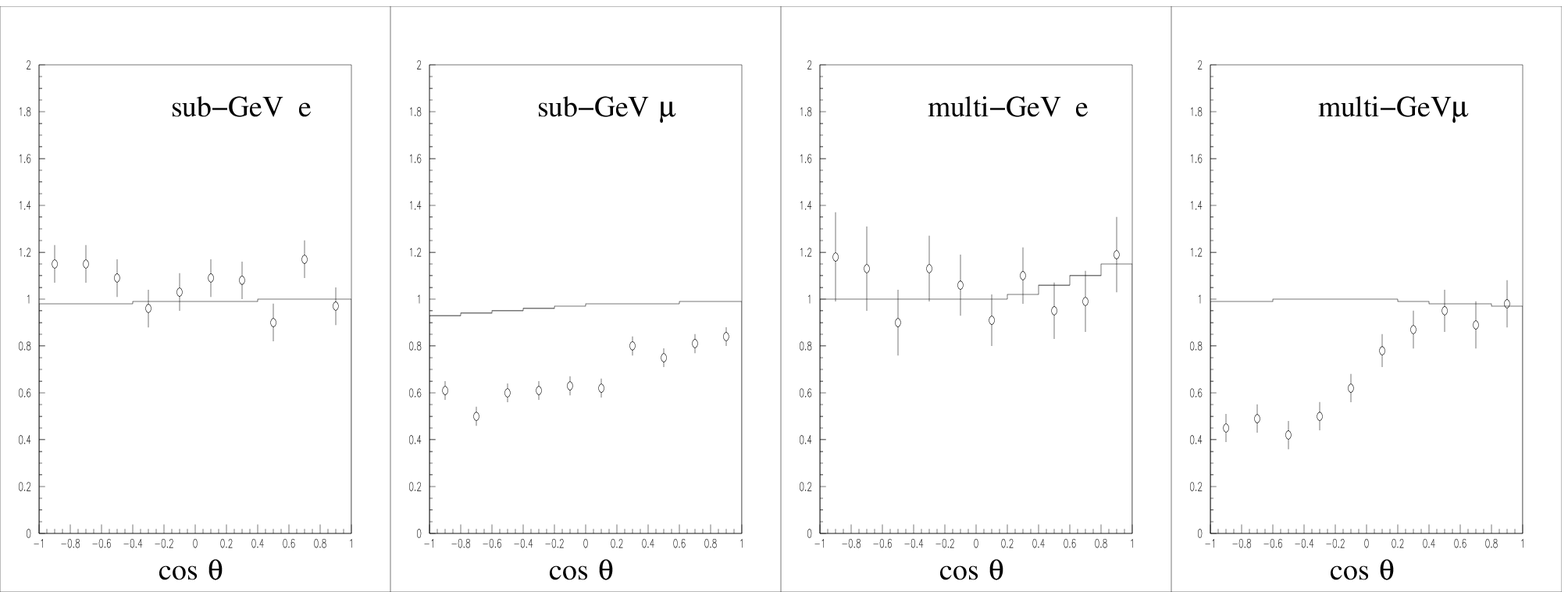} \hfill
\includegraphics[width=9cm]{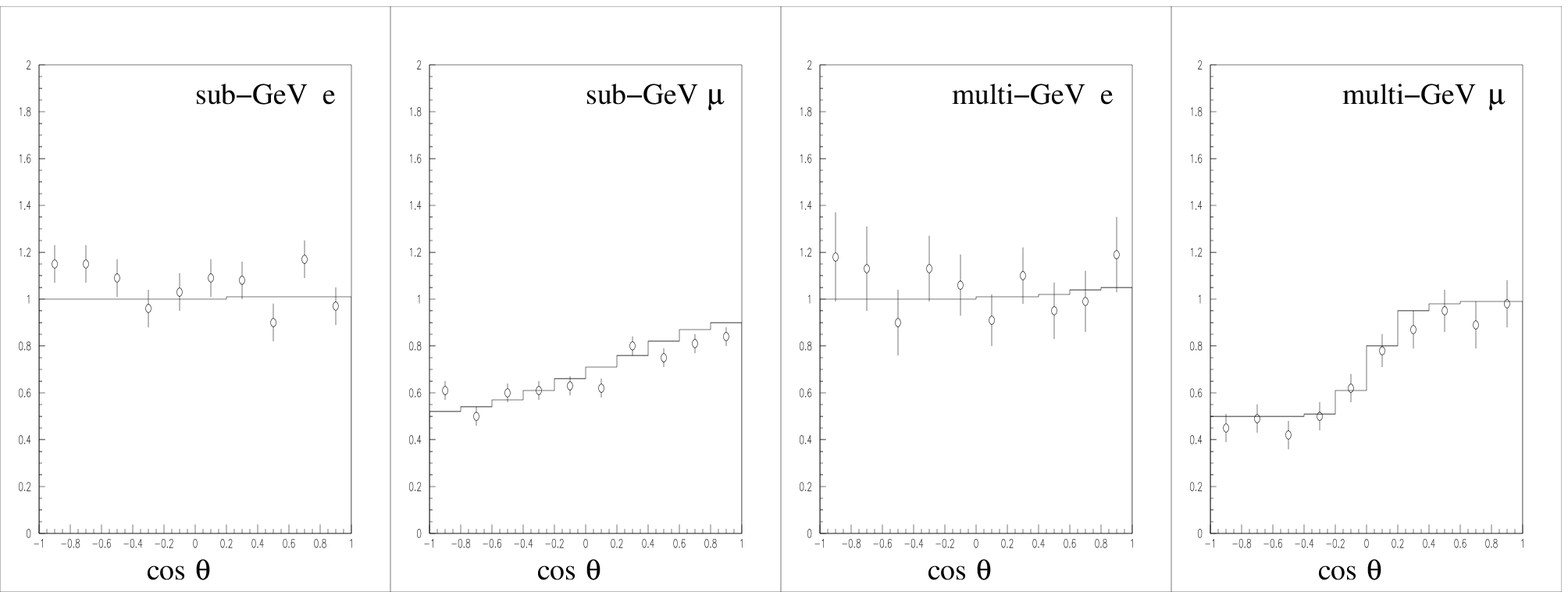}\hfill
\includegraphics[width=9cm]{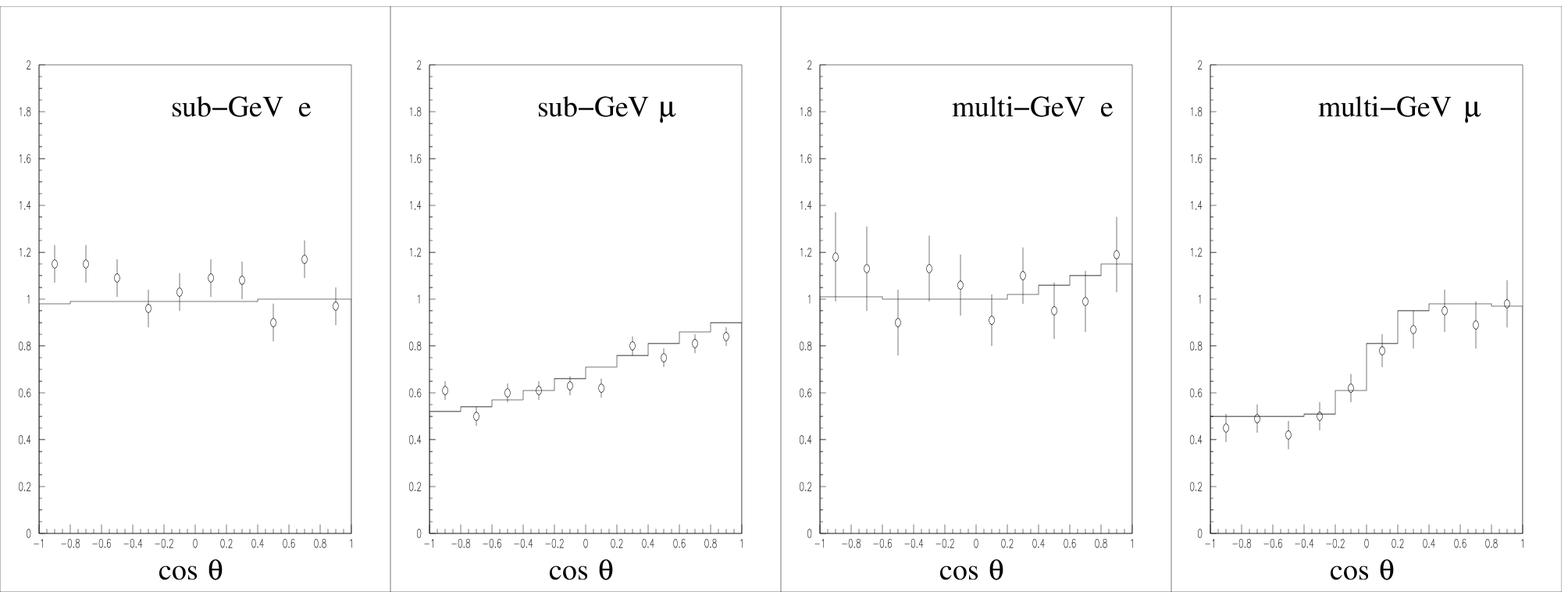}
\caption{Decoherence fits, from top to bottom: (a) pure decoherence in 
antineutrino sector, (b) pure decoherence in both sectors, (c) mixing plus
decoherence in the antineutrino sector only, (d) mixing plus decoherence
in both sectors. The dots correspond to SK data.}
\label{bestfit}
\end{figure}

As bare eye comparisons can be misleading, we have also calculated the
$\chi^2$  value for each of the cases, defining the atmospheric 
$\chi^2$  as
\begin{equation}
  \chi^2_{\rm atm}= \sum_{M, S}\sum_{\alpha=e,\mu}\sum_{i=1}^{10} 
     \frac{(R_{\alpha,i}^{\rm exp}-
     R_{\alpha,i}^{\rm th})^2}{\sigma_{\alpha i}^2} \quad .
\end{equation}
Here $\sigma_{\alpha,i}$ are the statistical errors,
the ratios $R_{\alpha,i}$ between the observed and predicted signal 
can be written as
\begin{equation}
  R_{\alpha,i}^{\rm exp}= N_{\alpha,i}^{\rm exp}/N_{\alpha,i}^{\rm MC}
\end{equation}
(with $\alpha$ indicating the lepton flavor and $i$ counting the
different bins, ten in total)
and $M,S$ stand for the multi-GeV and sub-GeV data respectively.
 For the CHOOZ experiment we used the 15 data points with their
statistical errors, where in each bin we averaged the probability
over energy and for LSND one datum has been included.
The results  with which we hope  all our claims 
become crystal clear are summarized in Table 1 and 2 , were we
present the $\chi^2$ comparison for the following cases:
(a) pure decoherence in the antineutrino sector, (b) pure
decoherence in both sectors, (c) mixing plus decoherence in
the antineutrino sector, (d) mixing plus decoherence in
both sectors, and (e) mixing only - the standard scenario
for all experimental data with (Table 2) and without (Table 1)
KamLAND spectral distortion .

\begin{table}[h]
\centering
\begin{tabular}{|c|c|c|}
\hline\hline
model & $\chi^2$ without LSND  & $\chi^2$ including LSND  \\ [0.5 ex]
\hline\hline
(a) & 1097.6 & 1104.3\\ \hline
(b) & 1037.8 & 1044.4\\ \hline
(c) & 45.7& 45.9 \\\hline
(d) & 52.5 & 52.7 \\\hline
(e) & 53.9 & 60.7  \\[1ex]
\hline\hline
\end{tabular}
\caption{$\chi^2$ obtained for (a) pure decoherence in 
antineutrino sector, (b) pure decoherence in both sectors, (c) mixing plus
decoherence in the antineutrino sector only, (d) mixing plus decoherence
in both sectors, (e) standard scenario with and without the LSND result 
for all experimental data but KamLAND spectral distortions.}
\end{table}

\begin{table}[h]
\centering
\begin{tabular}{|c|c|c|}
\hline\hline
model & $\chi^2$ without LSND  & $\chi^2$ including LSND  \\ [0.5 ex]
\hline\hline
(a) & 1135.8 & 1142.3\\ \hline
(b) & 1095.6 & 1102.3\\ \hline
(c) & 83.9 & 84.2 \\\hline
(d) & 90.7 & 90.9 \\\hline
(e) & 78.6 & 85.4  \\[1ex]
\hline\hline
\end{tabular}
\caption{$\chi^2$ obtained for (a) pure decoherence in 
antineutrino sector, (b) pure decoherence in both sectors, (c) mixing plus
decoherence in the antineutrino sector only, (d) mixing plus decoherence
in both sectors, (e) standard scenario with and without the LSND result, for all data. }
\end{table}

{}From the tables it becomes clear that the mixing plus decoherence scenario
in the antineutrino sector can easily account for all the available 
experimental information,
including LSND when KamLAND spectral distortion data are excluded but
does not significantly improve over the standard case once they are
included, showing clearly the limitations of our simplified scenario.

Such a change can be easily undestood by noticing that scenario (c)
improves the fit over the standard case (case (e)) by $\Delta \chi^2 =14.8 $ 
(see Table 1) but the price to pay for it is to give up spectral distortions 
in KamLAND.  In fact, for KamLAND data alone we find a $\chi^2 = 24.7 $
for case (e) and  $\chi^2 = 38.4 $ for an energy independent suppression
(case (c)), implying a $\Delta \chi^2 =13.7 $, which essentialy means that
the net effect is to put our decoherence model back to the standad case
level.

It is important to stress once more that our sample
point was not obtained through a scan over all the parameter space,
but  by an educated guess, and therefore plenty of room is left
for improvements. On the other hand, for the mixing-only/no-decoherence 
scenario,
we have taken the best fit values of the state of the art analysis
and therefore no significant improvements are expected. 
At this point a word of warning is in order: although superficially
it seems that scenario (d), decoherence plus mixing in both sectors,
provides an equally good fit, one should remember that including
decoherence effects in the neutrino sector can have undesirable
effects in solar neutrinos, especially due to 
the fact that decoherence
effects are weighted by the distance traveled by the neutrino, 
something that may lead to seizable (not observed!) effects in the
solar case.

Another point to stress is that, as can be easily seen in
the figure, decoherence plays no role  for atmospheric neutrino
energies and baselines, where 
the effect is given almost totally by the oscillations driven
by the mixing. This 
was somehow expected, since decoherence is known not to 
be the leading force behind the atmospheric neutrinos. 
In fact we are taking advantage
of the fact that for our particular set of parameters, no significant
effects are expected in the (2,3) channels. 

One might wonder then, whether decohering effects, which affect
the antineutrino sector sufficiently to account for the LSND result, 
have any impact on the solar-neutrino 
related parameters, measured through 
antineutrinos in the KamLAND experiment \cite{kamland,kamspec}. 
In order to answer this question,
 it will be sufficient to calculate the electron survival probability
 for KamLAND in our model, which turns out to be 
$ P_{\bar\nu_{\alpha}\rightarrow \bar\nu_{\beta}} \mid_{\mbox{\tiny  KamLAND}}
\simeq .57$, in perfect agreement with observations. As is well known, 
KamLAND is sensitive to a bunch of different 
reactors with distances spanning from 80 to 800 km. However, the bulk of the
signal comes from just two of those, whose distances are 160 and 179 km. These
parameters have been used to compute the survival probability.
Nevertheless, decohering effects, at least of the 
simplified type discussed 
in this work, 
are not able to account for any
spectral distortion; they rather provide an overall suppression. Thus,
the spectral distortion that is apparently being observed by KamLAND~\cite{kamspec},
tends to favor the standard three-generation scenario (without KamLAND
spectral information but including LSND the $\chi^2$ for models (c) and
(e) would be 45.9 and 60.7, respectively). Therefore, if KamLAND evidence
gets stronger  a simplified decoherence model of the class 
discussed here would be ruled-out.However, this should by no means 
be regarded as implying 
that decoherence models in general will not survive such a case.
On the contrary, it is our belief that CPT Violating decoherence models
with complex energy dependences of the decoherence 
parameters as the one discussed here, but probably of more complicated,
even non linear form, stand a good chance of describing nature.

It is also interesting
to notice that in our model,
  the LSND effect is not given  by the phase inside
the oscillation term ( which is proportional to the mass difference)
 but rather by the decoherence factor multiplying the oscillation term.
 Therefore the tension between LSND and
KARMEN  \cite{karmen} data is naturally eliminated, because the difference
in length leads to an exponential suppression. Thus, while we predict
a 0.24 \% anti-electron neutrino appearance probability for LSND, 
the corresponding one for KARMEN
gets only to 0.14\%. And although this number is already 
below their experimental
sensitivity, one should notice that KARMEN combines neutrino and antineutrino
channels in their analysis, so that the actual appearance probability is 
even smaller, as our effect shows up only in the antineutrino sector. 

Another potential source of concern for the present model of 
decoherence 
might be accelerator neutrino
experiments, which involve high energies and long baselines, and where
the decoherence $L \cdot E$ scaling can potentially be probed. 
This, however, 
is not the case. Accelerator experiments typically join their neutrino and
antineutrino data, with the antineutrino statistics being always smaller than
the neutrino one. This fact, together with the smaller antineutrino cross
section, renders our potential signal consistent with the background
contamination. Even more,  in order to constrain decoherence
effects of the kind we are proposing here through accelerator experiments,
excellent control and knowledge of the beam background are mandatory. 
The new KTeV  data \cite{kk} on kaon decay branching 
ratios, for example, will change the $\nu_e$ background enough to make any 
conclusion on the viability of decoherence models useless. After all, the 
predicted signal  in our decoherence scenario will be at the level of the 
electron neutrino contamination, and therefore one would need to disentangle 
one from the other.

Having said that,
it is clear that althougth this very simplified decoherence model 
(once neutrino mixing is taken into account) does not account
for all the observations including 
the LSND result better than the standard three-neutrino model, 
it certainly suggests 
that less simplified models are worth exploring. This scenario , 
which makes dramatic predictions for
the upcoming neutrino experiments, expresses a strong observable form of 
CPT violation in the  
laboratory, and, in this sense, our fit does not seem to give (yet) a 
conclusive answer to the 
question asked in the introduction as to whether the weak form of CPT 
invariance (\ref{probs}) is violated in Nature, but it certainly 
encourages further studies. 

This CPT violating pattern, with equal mass spectra for
neutrinos and antineutrinos, will have dramatic signatures in 
future neutrino oscillation experiments. The most striking 
consequence
will be seen in MiniBooNE~\cite{miniboone},
According to our picture, MiniBooNE will be able to confirm LSND
only when running in
the antineutrino mode and not in the neutrino one, as decoherence
effects live only in the former. Smaller but experimentally 
accessible signatures
will be seen also in MINOS~\cite{minos}, by comparing conjugated channels (most
noticeably, the muon survival probability).

We next remark that fits with decoherence
parameters with energy dependences of the form (\ref{decohparam}) 
imply that the 
exponential factors $e^{\lambda_k t}$ in (\ref{trans})
due to decoherence will  modify the amplitudes of the oscillatory terms
due to mass differences, and while one term  depends on  
$L/E$ the other one is driven by $L\cdot E$, where
we have set $t=L$, with $L$ the oscillation 
length (we are working with natural units where $c=1$). 

The order of the coefficients of these quantities, 
$\gamma^0_j  \sim 10^{-18}, 10^{-24}$~(GeV)$^2$, found in 
our sample point, implies that for energies of a few GeV,
which are typical of
the pertinent experiments, 
such values are not far from  
$\gamma_j^0 \sim \Delta m_{ij}^2$. If our conclusions
survive the next round of experiments, and therefore if 
MiniBOONE experiment~\cite{miniboone} confirms previous LSND claims,
then this may be a significant result. One would be tempted to 
conclude that if the above estimate holds,  
this would probably mean that 
the neutrino mass differences
might 
be due to quantum gravity decoherence. 
Theoretically it is still 
unknown how the neutrinos acquire a mass, or what kind of mass 
(Majorana
or Dirac) they possess. There are scenaria in which the mass of neutrino 
may be due to some peculiar backgrounds of string theory for instance.
If the above model turns out to be right we might then have, 
for the first time in low energy physics, 
an indication of a direct detection of a quantum gravity effect, which 
disguised itself as an induced decohering neutrino mass difference. 
Notice that in our  sample point only antineutrinos have non-trivial 
decoherence parameters $\overline{\gamma_{i}}$ , for $i=1$ and 3,
while the corresponding quantities in the neutrino sector vanish.  
This implies that there is a single cause for mass differences,
the decoherence in antineutrino sector, which is compatible 
with common mass differences in both  sectors. 
If it turns out to be true, this would be truly amazing. 

\section{Speculations on Theoretical Models of Decoherence}

Moreover, if the neutrino masses
are actually related to decoherence as a result of quantum gravity, this may
have far reaching consequences for our understanding of the Early
stages of our Universe, and even the issue of Dark Energy that came up
recently as a result of astrophysical observations
on a current acceleration of the Universe from either distant supernovae
data~\cite{sn} or measurements on Cosmic Microwave Background 
temperature fluctuations from the WMAP satellite~\cite{wmap}.   
Indeed, 
as discussed in ~\cite{emn,lopez}, 
decoherence implies an absence of 
a well-defined scattering S-matrix, 
which in turn would imply CPT violation in the 
strong form, according to the theorem of \cite{wald}. 
A positive cosmological 
{\it constant} $\Lambda > 0$ will also lead 
to an ill definition of an S-matrix, precisely due to
the existence, in such a case, 
of an asymptotic-future de Sitter 
(inflationary) phase of the universe, 
with Hubble parameter $\sim\sqrt{\Lambda}$, implying the existence of a 
cosmic (Hubble) horizon. This in turn will prevent a proper definition
of pure asymptotic states. 

We would like to point out at this stage 
that the claimed value of the dark energy density component
of the (four-dimensional) Universe today, $\Lambda \sim 10^{-122}M_P^4$,
with $M_P \sim 10^{19}$~GeV (the Planck mass scale),  
can actually be accounted for (in an amusing coincidence?) 
by the scale of the neutrino mass differences
used in order to explain the oscillation experiments.
Indeed, $\Lambda \sim [(\Delta m^2)^2/M_P^4]M_P^4 \sim 10^{-122} M_P^4$ for 
$\Delta m^2 \sim 10^{-5}$~eV$^2$, the order of magnitude  
of the solar neutrino mass difference assumed in oscillation 
experiments (which is the one that encompasses the decoherence effects,
as can be seen by eq.(\ref{prob})). 
The quantum decoherence origin of this mass then would 
be in perfect agreement with the decoherence properties of the cosmological
constant vacuum, mentioned previously. 
We note at this stage that it is possible to justify theoretically 
the above form for $\Lambda \propto (\Delta m^2)^2$, if one computes
the vacuum expectation value 
of the stress tensor of the neutrino field in a Robertson-Walker 
space-time background, not with respect to the mass-eigenstate vacuum, 
but with respect
to a ``flavour vacuum'', which may not be unitarily equivalent to the former.
Details are given in \cite{bm2} and references therein, where we refer
the interested reader. We only note here 
that the issue is related to the old problem of mixing 
in a quantum field theory, which has been studied in detail in \cite{vitiello}.
Needless to say that we by no means consider such a problem as 
completely understood to date, 
but we would simply like to point out here, and in \cite{bm2}, yet another
interesting feature, namely a possible link of the neutrino mass difference
to a non zero, non perturbative, contribution to the cosmological constant. 

Unfortunately, at present we have no concrete theoretical 
model to explain the above-described decoherence in the antineutrino
sector, with the non universal energy dependence (\ref{decohparam}). 
Conventional quantum-gravity decoherence scenaria assume usually a 
universal CPT conserving character for the decoherence parameters
between the particle and antiparticle sectors, and in fact they
yield too small decoherence parameters to be detected experimentally. 
For instance, in the two-level double-commutator model of  
quantum-gravity induced decoherence of \cite{adler},
the double commutator structure in the decoherence part yields
the estimate: $\gamma \sim (\Delta E)^2/M_P$ for a four-dimensional
quantum-gravity decoherence model. Such an estimate stems from the 
double commutator of the Lindblad operators $[b_i, b_i, \rho]$
when considered between appropriate energy eigenstates. 
Here, $\Delta E$ is the 
energy variance, which in the case of neutrinos may be estimated 
from the Hamiltonian $p + m^2/2p$ to be $\Delta E \sim (\Delta m^2)/E$ ,
thereby yielding a decoherence model with $\gamma \propto 
(\Delta m^2)^2/M_P E^2$, 
which is universal between particle and antiparticle sectors, 
and actually too small to be detected in oscillation experiments. 

It is possible, however, that 
other models yield significantly larger values to match 
our estimates for the decoherence parameters above.        
For instance, let us consider a crude model of 
decoherence, based on non-critical string theory~\cite{emn},
according to which a modified Lindblad evolution 
of the type (\ref{lindblad}) can be derived as a result of departure from
world-sheet conformal invariance. The latter describes the interaction of 
string matter with `fuzzy' space-time singular backgrounds, entailing
`loss of information'.
Using for the moment some technical jargon, we mention that 
departure from conformal invariance 
is formally manifested through the appearance of a central charge
deficit in the corresponding $\sigma$ model~\cite{emn}, 
which from 
a target-space viewpoint may resemble 
a cosmological constant $\Lambda$. 
Universes with a cosmological constant
are not conformal backgrounds of string theory, as a result of the 
asymptotic future Hubble horizon. 
In $\sigma$-models 
describing the interaction of 
matter with foamy backgrounds such $\Lambda$ does not have to 
be as small as the actual cosmological constant of our Universe, 
mentioned above. 
In fact it could be due to string loop corrections, 
that is higher world-sheet topologies, 
and this   
may lead to 
violations of conformal invariance of order of the string scale $M_s$. 

It has been argued  in \cite{mlambda} that such 
models of non critical 
string theory lead to a Lindblad type 
modified evolution for reduced density matrices
of four-dimensional string matter of the form:
\begin{equation} 
\partial_t \rho = i[\rho, H_{\rm eff}] + {\cal L} :g_{\mu\nu} 
[g^{\mu\nu} , \rho] :~,
~\quad 
{\cal L} = {\cal O}(M_s \Lambda)
\label{liouv} 
\end{equation} 
where $M_s$ is the string scale, which, in general, may be  
different 
from the four-dimensional
Planck mass $M_P\sim 10^{19}$~GeV, 
$g^{\mu\nu}$ is the (fuzzy) metric background in ten dimensions
(the critical dimension of string theory), the : \dots : indicate
quantum ordering of the appropriate operators, 
and $\Lambda$ denotes now 
the ten-dimensional dimensionless central-charge
deficit,
which, if due to string loop corrections, it could be of order one
(in natural string ($M_s$) units). 
One may select an antisymmetric quantum ordering prescription,
which leads to a double commutator structure of the decoherence
terms ${\cal L} :[g_{\mu\nu}, [g^{\mu\nu} , \rho]] :$. 
This will be 
a local contribution to matter propagation, due to 
interactions of matter with space time foam,
and thus will be dominant as compared to global contributions
due to a non-zero Dark energy component of the Universe~\cite{mlambda}.
It remains a challenge of course to explain how such a local 
decoherence
will not lead to large global contributions to the vacuum energy. 

The decoherence terms in (\ref{liouv}) are essentially proportional
to the world-sheet $\beta$ function for the graviton in a `cosmological 
constant' space time  with metric $g_{\mu\nu}$, 
i.e. of the form $\beta_{\mu\nu} = M_s^{-2}
R_{\mu\nu} \sim \Lambda g_{\mu\nu}$~\cite{mlambda},
where $R_{\mu\nu}$ is the Ricci tensor of the cosmological constant
ten-dimensional non-conformal string background (an ${\cal O}(\alpha ' 
= M_s^{-2})$ analysis for the world-sheet 
$\beta$-functions of the pertinent $\sigma$-model suffices for our purposes).

If we accept this scenario for time evolution of matter (or antimatter), 
then, in our case of a non-zero string-loop induced
cosmological constant, 
upon projecting on string theory space states $|g^i>$,
where $g^i$ are background fields in the target space
over which the matter string propagates (such as $g_{\mu\nu}$ etc.),
the double commutator structure 
yields simply~\cite{emn} the square of the variance of the metric background
as a result of interaction with stringy matter
$(\Delta g_{\mu\nu})^2$. This expresses essentially the 
back reaction of matter onto space time, as a result of the quantum 
properties of the foam. 

The form of this variance depends on the details of the model, 
and it is conceivable, although we have no concrete model 
in mind at this stage, that the variances are non zero only 
for the antineutrino sector, and such that 
they allow
to write 
(\ref{liouv}) in the form 
(\ref{expandedlind}), where the Lindblad operator is diagonal,
with various entries that could match our estimates above. 
In the case of antineutrino decoherence, for instance, 
one may associate the variance with 
the uncertainty in the measurement of a typical length in the 
problem, which is the oscillation length. In our 
model the latter is of order
$L \sim E/\Delta m^2$ in both particle and antiparticle sectors.
Thus, an uncertainty in $\Delta g_{\mu\nu} $ may be associated
with an uncertainty in the covariant length~\cite{ng} 
$\Delta L^2 \sim L^2 \Delta g_{\mu\nu}$, from which we
obtain the estimate: $(\Delta g_{\mu\nu} )^2 \sim (\Delta L^2 /L^2)^2 \sim 
(\delta E /E)^2$. Here, $\delta E$ denotes an antineutrino, say,  
energy uncertainty due to the `fuzziness' of space time,  and should not be 
confused with the energy variance between neutrino states mentioned 
above in the context of the model of \cite{adler}. 

It is such uncertainties that could be energy dependent in a peculiar
way only for the antineutrino sector, as a result of the strong form 
of CPT violation due to decoherence~\cite{wald},  
e.g. $\delta E \sim E (E/M_s)^\alpha$, for some $\alpha$
which are in general different for the different decoherence
parameters,  
and yield the complicated energy dependence argued in our work
above. In fact, if the conformal invariance violations in (\ref{liouv}) 
are large only in the antineutrino sector, then this may provide 
the natural explanation as to why the dark energy component 
of the Universe is still small, given that the latter will be 
due to 
matter, and there is a huge matter-antimatter asymmetry in the Universe. 
At present, however, we have no answers to such issues, and our remarks
above should be treated with caution. 
We find it challenging, nevertheless, 
to try and interpret theoretically, in terms of microscopic models, 
the complicated energy dependence of our decoherence parameters
(\ref{decohparam}). This is currently under progress.

In any case, given that there is no  full/complete theory of decoherence,
any guess on the order of magnitude of the coefficients may be proven wrong.
This fact encourages phenomenological analyses like ours, whose results
can be contrasted to experimental data and may remain valid independently
of the theoretical speculations on the origin and size of the $\gamma$'s.
 
\section{Instead of Conclusions} 

In this work we have presented a set of decoherence parameters
which can account for  all available neutrino data, including LSND results.
By invoking a three-generation decoherence scenario, combined with 
conventional oscillations, and flavor violations, we have managed to 
fit the available data and to obtain decoherence parameters which
could find a natural explanation in some theoretical  
models of quantum gravity. We explained the LSND results
not by assuming CPT violating mass differences, but by invoking 
CPT violation in its strong form, namely that the decoherence
parameters differ between neutrinos and antineutrinos. 
The agreement of this fit with the data was found to be better
than in the two-generation decoherence case~\cite{lisi}, where there was
no evidence for decoherence. This conclusion depends crucially upon
the inclusion of the LSND results. 
Our analysis and conclusions are therefore in disagreement with pessimistic
claims made in the recent literature~\cite{lisi2,gago} 
on the non-necessity 
of studies of quantum-gravity-induced decoherence in three generation models.

It is imperative that the LSND results are confirmed by future
experiments, such as MiniBOONE~\cite{miniboone}, 
because as we have 
seen above, it is indeed possible that the study of neutrino
physics constitutes a true
and rare window to Planck scale physics, with far reaching consequences
ranging from cosmological considerations to the origin of the dark energy
component of the Universe. 
In the present work, motivated by
our sample point, we put forward a conjecture 
that quantum-gravity decoherence may be responsible for the neutrino
mass differences and the non-zero value of the vacuum energy density
of the Universe today. The scaling of the decoherence parameter
with the energy is a crucial test of such conjectures, and 
would be a useful theoretical guide for model building, should
the results of LSND and the present work survive future data. 

Before closing we must stress the fact that the dynamical semi-groups
approach to decoherence, examined here, relies on the approximation that 
the quantum gravity effects  imply a linear decoherence, i.e. a linear 
dependence on the reduced density matrix of matter. This may not be true
in a complete theory of quantum gravity, and hence, even if we get agreement
with experimental data, this agreement should be interpreted with great care.
Clearly we are in need of a complete and detailed mathematical
model of quantum gravity, because generic and model independent analyses
may be misleading. Nevertheless, the possibility of verifying 
experimentally deviations from conventional field theoretical treatments
in the near future, e.g. in the context of neutrino physics, is by itself
exciting even if it cannot give a complete answer to such fundamental
questions as the Planck scale structure of quantum space time. 

\section*{Acknowledgments} 

We thank Bill Louis, Joe Lykken and Chris Quigg for comments and suggestions and
Janet Conrad, Graciela Gelimini, Silvia Pascoli and  Mike Shaevitz for valuable
input.
The work of of GB is supported by CICYT, Spain,
under grant FPA2002-00612, while that 
of NEM is partly supported by the European Union 
(contract HPRN-CT-2000-00152).

\end{document}